\documentclass[10pt,twocolumn,showpacs,floatfix,prl]{revtex4}
\usepackage[latin9]{inputenc}
\usepackage{color}
\usepackage{amsmath}
\usepackage{amssymb}
\usepackage{graphicx}
\usepackage{esint}

\makeatletter
\@ifundefined{textcolor}{}
{%
 \definecolor{BLACK}{gray}{0}
 \definecolor{WHITE}{gray}{1}
 \definecolor{RED}{rgb}{1,0,0}
 \definecolor{GREEN}{rgb}{0,1,0}
 \definecolor{BLUE}{rgb}{0,0,1}
 \definecolor{CYAN}{cmyk}{1,0,0,0}
 \definecolor{MAGENTA}{cmyk}{0,1,0,0}
 \definecolor{YELLOW}{cmyk}{0,0,1,0}
}

\usepackage{epsf}
\usepackage{amsfonts}

\newcommand{\ve}{\vec e}

\newcommand{\vA}{\vec A}

\newcommand{\be}{\begin{equation}}
\newcommand{\ee}{\end{equation}}

\renewcommand{\vec}[1]{\mathbf{#1}}

\renewcommand{\vr}{\mathbf{r}}

\makeatother

\begin{document}

\title{The weak side of strong topological insulators}

\author{Bj\"orn Sbierski, Martin Schneider, and Piet W. Brouwer}

\affiliation{Dahlem Center for Complex Quantum Systems and Fachbereich Physik,
Freie Universit\"at Berlin, 14195 Berlin, Germany}

\date{\today}
\begin{abstract}
Strong topological insulators may have nonzero weak indices. The nonzero
weak indices allow for the existence of topologically protected helical
states along line defects of the lattice. If the lattice admits line
defects that connect opposite surfaces of a slab of such a ``weak-and-strong'' topological insulator, these states effectively connect
the surface states at opposite surfaces. Depending on the phases accumulated along the dislocation lines, this connection results in a suppression of in-plane transport and the opening of a spectral gap or in an enhanced density of states and an increased conductivity.
\end{abstract}
\maketitle

\emph{Introduction.---} Band insulators come in topologically distinct
classes, where the topologically nontrivial classes have extended
surface states, which are robust to small deformations of the Hamiltonian
\cite{kane2005,kane2005b,bernevig2006b,moore2007,fu2007,roy2009}.
The topological classification of generic band insulators in three
dimensions distinguishes ``strong'' and ``weak'' topological indices
\cite{fu2007,fu2007b}. A nonzero value of the strong
index signifies a ``strong topological insulator''; Surface states
of strong insulators have a spectrum with an odd number of Dirac cones,
and they are robust to disorder or other perturbations that break
the lattice translation symmetry. In a ``weak topological insulator'',
{\em i.e.}, if the strong invariant is trivial and the weak invariant 
is nontrivial, the lattice translation symmetry is 
essential for the protection of the surface states,
although, as was pointed out in a seminal article by Ringel
\emph{et al.} \cite{ringel2012}, the surface states of a weak
topological insulator
show a remarkable robustness in the presence of perturbations
that preserve the lattice translation symmetry on the average \cite{mong2012,fulga2014}.

An important property of insulators with nontrivial weak indices
is that a line
dislocation may have topologically protected helical states, similar to
the helical edge states of a two-dimensional topological insulator
\cite{teo2010,ran2009}. The precise conditions for the existence
of such strongly protected states depends on the Burgers vector $\mathbf{b}$
of the dislocation \cite{ran2009,imura2011}. The helical states along
the dislocation line remain topologically protected as long as the
notion of a separate dislocation with a well-defined Burgers vector
remains valid.
The presence of nonzero weak and strong indices is not mutually exclusive,
and it is possible that a band insulator is at the same time a weak topological
insulator {\em and} a strong topological insulator. Such a scenario is expected
to be relevant, \emph{e.g.}, for BiSb compounds or for the putative
Kondo topological insulator SmB$_{6}$ \cite{ando2013}. In principle,
such a ``weak-and-strong topological insulator'' combines an odd
number of Dirac cones in the surface-state spectrum with topologically
protected helical states along lattice defects.

\begin{figure}
\begin{centering}
\includegraphics{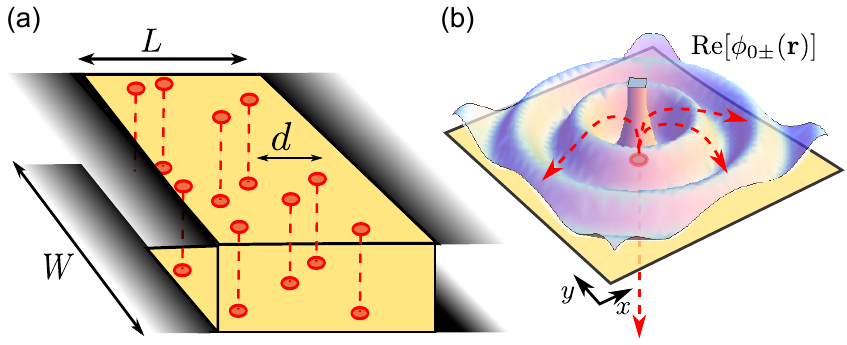}
\par\end{centering}

\protect\caption{\label{fig:setup}(a) Topological insulator slab of size $L\times W$,
with top and bottom surfaces connected via randomly placed dislocation
lines with mean distance $d$. Ideal contacts are attached to the
left and right, for top and bottom surfaces separately. (b) Zero-angular-momentum
($m=0$) radial waves for nonzero wavenumber $k$ at the surface of
the topological insulator are transmitted perfectly into and out of
the one-dimensional helical states along the dislocation line.}
\end{figure}

Realistic topological insulators are often layered materials, and
flakes of such materials are usually investigated in a quasi-two-dimensional slab geometry, in which the slab thickness is large
enough that surface states at the bottom and top surfaces remain well
separated. The presence of dislocation lines that connect the top and
bottom surfaces of a weak-and-strong topological insulator,
as shown schematically in Fig.\ \ref{fig:setup}(a),
may, however, provide a mechanism by which the two 
surfaces are coupled nevertheless. As we show here, a finite density of
dislocation lines may lead to the opening of a gap in the
surface-state spectrum of a slab and to a strong suppression of
electron transport parallel to the surface, although the precise
scenario depends on the phase that electrons accumulate along the
dislocation line. The possibility of a
coupling of surface states at bottom and top surfaces via dislocation
lines presents a ``weak side'' of topological insulators with
nontrivial strong and weak indices; it does not exist for strong topological
insulators with trivial weak indices, for which dislocation lines do
not carry protected helical states.
We now proceed with a description of our results.

\noindent 

\emph{Description of dislocation line defect in terms of a $\pi$-flux
line.}--- The weak indices $\nu_{j} = 0,1$, $j=1,2,3$,
of a topological insulator are defined with respect to a basis $(\mathbf{G}_{1},\mathbf{G}_{2},\mathbf{G}_{3})$
of reciprocal lattice vectors. Together they uniquely define a reciprocal
lattice vector $\mathbf{M}=(1/2)(\nu_{1}\mathbf{G}_{1}+\nu_{2}\mathbf{G}_{2}+\nu_{3}\mathbf{G}_{3})$
\cite{ran2009}. As shown by Ran, Zhang, and Vishwanath, a lattice
dislocation binds an odd number of helical modes if and only if its
Burgers vector $\mathbf{b}$ satisfies \cite{ran2009} 
\begin{equation}
  e^{i \mathbf{b}\cdot\mathbf{M}}= - 1. \label{eq:BM}
\end{equation}
In that case, there is an odd number of surface-state Dirac cones
within which electrons pick up a phase $\pi$ upon going around the
position $\mathbf{r}_{{\rm d}}$ at which the dislocation line pierces
the surface. The low-energy Dirac Hamiltonian for such surface states
is accordingly 
\begin{equation}
H=v(\mathbf{p}+e\mathbf{A}/c)\cdot\boldsymbol{\sigma},\label{eq:H}
\end{equation}
where $v$ is the surface-state velocity, $\mathbf{p}=(p_{x},p_{y})$,
$\boldsymbol{\sigma}=(\sigma_{x},\sigma_{y})$, and $\mathbf{A}(\mathbf{r})$
is the vector potential corresponding to a flux line with flux $hc/2e$
at position $\mathbf{r}_{{\rm d}}$, a ``$\pi$-flux''. Since the
total number of Dirac cones in the surface-state spectrum is odd if
the strong index $\nu_{0}=1$, the number of surface cones described
by a Dirac Hamiltonian without $\pi$-flux line is even if $\mathbf{b}\cdot\mathbf{M}$
is an odd multiple of $\pi$ \cite{ran2009}. For simplicity we focus
on the minimal model, in which there is a single surface state with
low-energy effective Hamiltonian (\ref{eq:H}) in the vicinity of
a dislocation line for which the condition (\ref{eq:BM}) holds.

To elucidate the relation between the surface states and the helical
states propagating along the dislocation line, it is instructive to
analyze the eigenstates of the Hamiltonian (\ref{eq:H}) at energy
$\varepsilon=\hbar vk$ using polar coordinates $(r,\theta)$. We
choose the $\pi$-flux line --- the location where the dislocation line
pierces the surface --- as the origin. This is a problem that previously
was considered in the context of graphene \cite{heinl2013,schneider2014}.
With the choice $\mathbf{A}=(\hbar/2 er)\ve_{\theta}$,
where $\ve_{\theta}$ is the unit vector for the azimuthal
angle, the Hamiltonian (\ref{eq:H}) is invariant under rotations,
so that we can look for eigenstates of the total angular momentum
$j_{z}=l_{z}+(\hbar/2)\sigma_{z}$. These have the form 
\begin{equation}
\psi_{m}(\mathbf{r})=\left(\begin{array}{c}
e^{i(m-1)\theta}\phi_{m+}(r)\\
e^{im\theta}\phi_{m-}(r)
\end{array}\right),\label{eq:psi_m}
\end{equation}
where $m$ is an integer and the radial wavefunctions $\phi_{m\pm}(r)$
satisfy 
\begin{eqnarray}
\left(\partial_{r}\pm\frac{m\pm1/2}{r}\right)\phi_{m\mp}(r) & = & ik \phi_{m\pm}(r).\label{eq:phi}
\end{eqnarray}
For generic $m$ there is a single regular solution of Eq.\ (\ref{eq:phi}), which describes the scattering of radial waves
off the flux line. An exception is the case $m=0$, for which there
are {\em two} linearly independent solutions 
\begin{equation}
\phi_{0\pm}(r)=\alpha_{{\rm out}}\frac{e^{ikr}}{\sqrt{r}}\pm\alpha_{{\rm in}}\frac{e^{-ikr}}{\sqrt{r}},
  \label{eq:m0}
\end{equation}
for which the amplitudes $\alpha_{{\rm out}}$ and $\alpha_{{\rm in}}$
of outgoing and incoming radial waves can be freely chosen. Since
time-reversal symmetry rules out backscattering for the $m=0$ states
and for the helical states propagating along the defect line %
\footnote{For the gauge chosen here, time-reversal amounts to the operation
$\psi\to i\sigma_{y}\psi^{*}$, followed by a gauge transformation
$\psi\to e^{i\theta}\psi$. This corresponds to the change
$m\to-m$, so that time-reversal symmetry forbids backscattering for
the $m=0$ mode only.%
}, the incoming $m=0$ mode must be fully transmitted into the outgoing
defect state, and the incoming defect state is fully transmitted into
the outgoing $m=0$ mode, as shown schematically in Fig.\ \ref{fig:setup}(b).

\emph{Surface states in the presence of dislocation lines.---}
We now consider transport properties and density of states of
surface states for a slab geometry with multiple
dislocation lines, piercing the top and bottom surfaces at random
positions, see Fig.\ \ref{fig:setup}(a). We choose coordinates such
that the bottom and top surfaces are parallel to the $xy$ plane,
with transport taking place in the $x$ direction. For simplicity
we take the dislocation lines to pierce bottom and top surfaces at
the same in-plane position $\mathbf{r}_{{\rm d}}=(x_{{\rm d}},y_{{\rm d}})$,
an assumption that is appropriate for a low-energy, long-wavelength
description of a thin slab. The in-plane dimensions of the slab are
$L\times W$, and we assume that the slab thickness is sufficient large,
so that surface states at the bottom and top surfaces do not overlap
in the absence of lattice dislocations. We take periodic boundary
conditions in the $y$ direction, choosing the aspect ratio $W/L$
large enough that the results of our calculation do not
depend on this choice of boundary conditions.

We calculate the density of states and the transport properties of 
the surface states using a scattering approach. 
%
%
The scattering matrix $S_{\sigma,\sigma'}$ links the amplitudes of incoming and outgoing waves in an ``ideal'' part of the two surfaces, to the left and right of a section with a finite density of dislocation lines. The indices $\sigma,\sigma^{\prime}=+1$, $-1$ for the top and bottom surface, respectively. Dislocation lines connect
the top and bottom surfaces, so that in general $S_{\sigma,\sigma^{\prime}}$
is not block diagonal. We denote the amplitudes of incoming and outgoing
waves to the left (right) of the section by 
vectors $a_{{\rm L}\sigma n}^{{\rm in}}$
and $a_{{\rm L}\sigma n}^{{\rm out}}$ ($a_{{\rm R}\sigma n}^{{\rm in}}$
and $a_{{\rm R}\sigma n}^{{\rm out}}$), respectively, where the index
$n$ refers to the transverse momentum $q_{n}=2\pi n/W$. With this
notation, the scattering matrix $S_{\sigma,\sigma^{\prime}}$ relates
outgoing and incoming waves as 
\begin{equation}
\left(\begin{array}{c}
a_{{\rm L}\sigma}^{{\rm out}}\\
a_{{\rm R}\sigma}^{{\rm out}}
\end{array}\right)=\sum_{\sigma^{\prime}=\pm}S_{\sigma,\sigma^{\prime}}\left(\begin{array}{c}
a_{{\rm L}\sigma^{\prime}}^{{\rm in}}\\
a_{{\rm R}\sigma^{\prime}}^{{\rm in}}
\end{array}\right).
\end{equation}
Each component $S_{\sigma,\sigma^{\prime}}$ can be decomposed into
transmission and reflection blocks in the standard way, 
\begin{equation}
S_{\sigma,\sigma^{\prime}}=\left(\begin{array}{cc}
r_{\sigma,\sigma^{\prime}} & t_{\sigma,\sigma^{\prime}}^{\prime}\\
t_{\sigma,\sigma^{\prime}} & r_{\sigma,\sigma^{\prime}}^{\prime}
\end{array}\right).\label{eq:Ssigmasigmap}
\end{equation}

\begin{figure}
\begin{centering}
\includegraphics{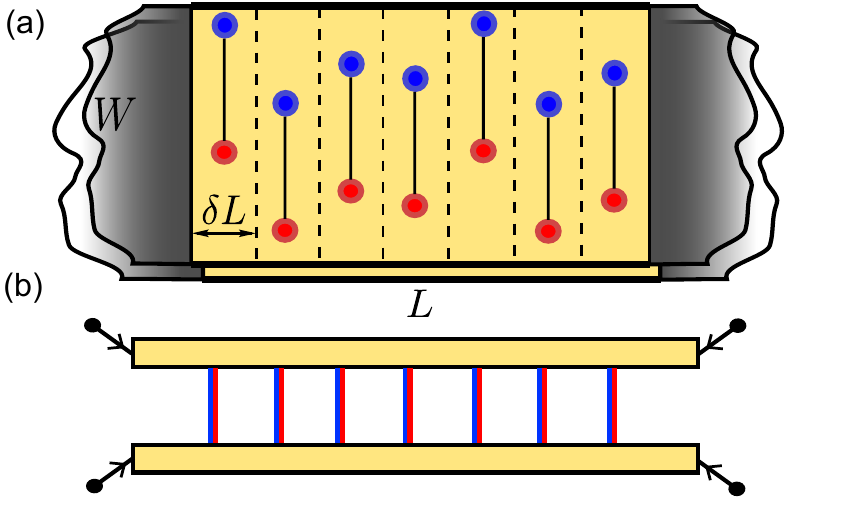} 
\par\end{centering}

\protect\caption{\label{fig:top-side-view}Schematic picture of a topview (a) and sideview
(b) of the topological insulator slab. The calculational scheme involves
the computation of the scattering matrix $\tilde{S}$ for a slab of
width $\delta L$, followed by the concatenation of scattering matrices
of individual slabs to obtain the scattering matrix $S$ of the full
structure. 
}
\end{figure}

Our strategy will be to first calculate the scattering matrix $\tilde{S}$ for a ``short'' slab of length $\delta L$ with only a pair of dislocation lines, and then calculate the scattering matrix $S$ of a slab of full length $L$ by concatenating scattering matrices of individual slices \cite{bardarson2007}, see Fig.\
\ref{fig:top-side-view} (top). We place a pair of dislocation lines at
$\vr_{{\rm d},1} = (x_{{\rm d}},y_{{\rm d},1})$ and $\vr_{{\rm d},2} = 
(x_{{\rm d}},y_{{\rm d},2})$, with
$0<x_{{\rm d}}< \delta L$ and $0<y_{{\rm d},1}<y_{{\rm d},2} < W$ randomly chosen.
Since the aspect ratio $W/L\gg1$, the pairwise placement of
dislocation in a slab (compared to placement of single dislocation
lines) does not affect the in-plane conductivity or the density of
states. It does, however, allow us to choose a gauge such that the
vector potential $\mathbf{A}(\mathbf{r})$ is nonzero for $x=x_{{\rm
d}}$ only,
\begin{equation}
\mathbf{A}(\mathbf{r})=\frac{h}{2 e} \delta(x-x_{{\rm d}})\ve_{x}\times\left\{ \begin{array}{ll}
1 & \mbox{if \ensuremath{y_{{\rm d},1}<y<y_{{\rm d},2}}},\\
0 & \mbox{else}.
\end{array}\right.\label{eq:Asharp}
\end{equation}
An important further parameter in the calculation is the phase shift $e^{i \theta_{\rm d}(\vr_{\rm d})}$ that electrons accumulate along the dislocation line.
For our calculations we found it advantageous to generalize the above
procedure to slabs with an even number $2n$ of dislocation lines.

The calculation of the scattering matrix $\tilde{S}$ for a slab with
a single pair of dislocation lines turned out to be an interesting
problem in its own right. Although the scattering problem for a single
dislocation line is easily solved in polar coordinates, see Eq.\ (\ref{eq:phi}),
we could not find a practical way to extract a scattering matrix for
the geometry of Fig.\ \ref{fig:setup}(a) from this solution. Instead,
we compute $\tilde{S}$ from a solution of the Dirac equation for
a regularized (\emph{i.e.}, smeared out) $\pi$-flux. (Without regularization
the scattering problem with a $\pi$-flux line cannot be solved numerically.)
The details of this calculation are given in the supplemental material
\footnote{See supplemental material for details.}.

\emph{Results.---} By concatenation of scattering matrices
for slices of length $\delta L$, each with an even
number $2n$ of dislocation lines, we can construct the full scattering
matrix $S$ for a slab of length $L$ with randomly placed dislocation
line pairs at concentration $1/d^{2}=N_{{\rm d}}/WL$, with $N_{{\rm d}}$
the total number of dislocation lines, see Fig. \ref{fig:top-side-view}(a).
The Landauer formula expresses the in-plane conductance $G_{\parallel}$ 
and the cross conductance $G_{\perp}$ in terms of the transmission and
reflection blocks of the scattering matrix $S$, 
\begin{eqnarray}
  G_{\parallel} &=& \frac{e^{2}}{h}\sum_{\sigma,\sigma'=\pm1}\mbox{tr}\, t_{\sigma,\sigma'}t_{\sigma,\sigma'}^{\dagger}, \ \
  G_{\perp} = \frac{e^{2}}{h} \mbox{tr}\,
  S_{+-} S_{+-}^{\dagger}.
\end{eqnarray}
For the calculation of the density of states, we consider a periodic array of slabs of length $L$. In this case the spectrum of Bloch states can be obtained from the condition that
\begin{equation}
  S_{\kappa_x}(\varepsilon) \equiv \left(\begin{array}{cc}
0 & e^{-i \kappa_x L}\\
e^{i \kappa_x L} & 0
\end{array}\right)S\left(\varepsilon\right)
\end{equation}
has a unit eigenvalue, where $\hbar \kappa_x$ is the crystal momentum.

\begin{figure}
\begin{centering}
\includegraphics{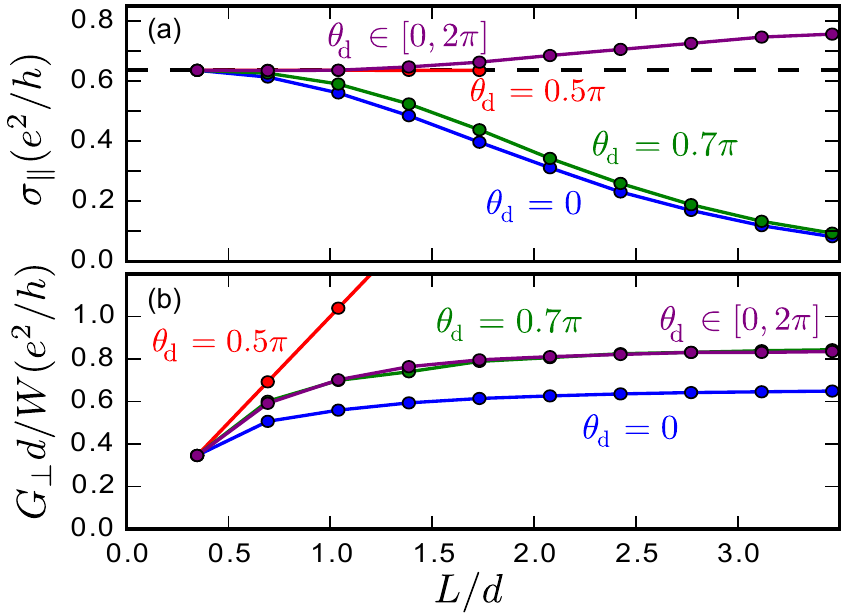}
\par\end{centering}

\protect\caption{\label{fig:results_transport}
Zero-energy in-plane conductivity $\sigma_{\parallel}$
(a) and cross conductance $G_{\perp}$ (b) 
for a slab of weak-and-strong topological insulator
with a concentration $1/d^{2}$ of randomly placed dislocation lines. 
The different curves refer to different choices for the phases 
$\theta_{{\rm d}}(\vr_{\rm d})$, as shown in the figure. 
The dashed line in (a) denotes the clean-limit in-plane conductivity 
$\sigma_{\parallel,0}=2 e^2/\pi h$.
Data points denote an average over 500 disorder realizations, statistical
error bars are typically smaller than the markers.}
\end{figure}

Results of the transport calculations are shown in Fig.\ \ref{fig:results_transport} for an average over 500 random realizations of the dislocation lines. The energy $\varepsilon$ is set to zero throughout the calculation, to maximize the effect of the dislocation lines. The sample length $L$ is measured in units of the mean distance $d$ between dislocation lines, which is the only fundamental length scale in the system at zero energy. The trivial $W$-dependence of $G_{\parallel}$ is eliminated by considering the in-plane conductivity $\sigma_{\parallel}=G_{\parallel}L/W$. For $L/d\to0$ we recover the clean-limit conductivity $\sigma_{\parallel,0}=2e^{2}/\pi h$ of a pair of decoupled topological-insulator surfaces \cite{katsnelson2006b,tworzydlo2006}.  Anticipating a proportionality $G_{\perp}\propto W/d$, in Fig.\ \ref{fig:results_transport}(b) we show $G_{\perp}d/W$ as a function of $L/d$. Unlike the longitudinal conductivity $\sigma_{\parallel}$, the cross conductance vanishes in the clean limit $L/d \to 0$.

We observe that the in-plane conductivity has a strong dependence on the phase that electrons pick up while traveling along the dislocation lines. In particular, if all phases are equal, $\theta_{\rm d}(\vr_{\rm d})=\theta_{\rm d}$ for all $\vr_{\rm d}$, $\sigma_{\parallel}$ is strongly suppressed for $L\gtrsim d$ except for $\theta_{\rm d} = \pm \pi/2$, for which we find that $\sigma_{\parallel}$ is independent of $L/d$ within numerical accuracy 
\footnote{In the case $\theta_{\rm d}=\pi/2$, we observe that the scattering matrix at zero energy ceases to be unitary for $L/d\gtrsim2$, which is the reason for the relatively small upper bound on the system sizes shown in Fig.\ \ref{fig:results_transport}. This can be understood from the perspective of bound state formation, once the number of bound states $N_{\mathrm{d}}$ is of the same order as the number of modes considered in the scattering matrix, $N_{\mathrm{eff}}$.}.
Figure \ref{fig:results_transport} shows the representative cases $\theta_\mathrm{d}=0$, $0.7\pi$, and $\pi/2$ and we also present the case $0 \, \leq \, \theta_{\rm d}(\vr_{\rm d}) \! < \! 2 \pi$ uniformly distributed, which shows a slight increase of $\sigma_{\parallel}$ with $L/d$. The $\theta_{\rm d}$-dependence of the cross-conductance is not as strong; $\theta_{\rm d}$ mainly determines the value at which $G_{\perp}$ saturates for large $L/d$. An exception is $\theta_{\rm d} = \pi/2$, for which we could not observe a saturation for the system sizes we could achieve.

\noindent 

\begin{figure}
\noindent \begin{centering}
\includegraphics{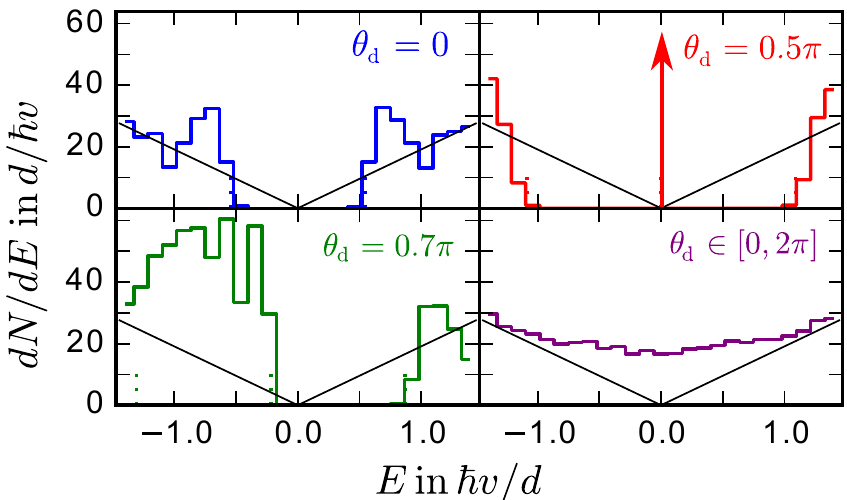}
\par\end{centering}

\protect\caption{\label{fig:DOS}Density of states $dN/dE$ of a sample with dislocation
line density $1/d^{2}$. The four curves represent the four representative scenarios for the choice of the phase shifts $\theta_{{\rm d}}(\vr_{\rm d})$, as explained in the text. The vertical dashed lines correspond to energies calculated from Eq.\ \eqref{eq:phase matching}. The thin black lines denote the ideal surface-state density of states without dislocation lines. The arrow represents a Dirac delta function at zero energy. Data points denote an average over $5$ disorder
realizations and $80$ values of the crystal momentum $\kappa_x$.}
\end{figure}

Results for the density of states are shown in Fig.\ \ref{fig:DOS}, again for four representative choices of the phase shifts $\theta_{\rm d}(\vr_{\rm d})$. For fixed $\theta_{\rm d}(\vr_{\rm d}) = 0$ we observe one or two gaps placed asymmetrically around $\varepsilon=0$. For generic fixed $\theta(\vr_{\rm d}) = \theta_{\rm d}$ (such as the case $\theta_{\rm d} = 0.7 \pi$ shown in the figure) we observe an asymmetric gap around $\varepsilon=0$. For $\theta_{\rm d}(\vr_{\rm d}) = \pi/2$ a symmetric gap is restored, but with one mid-gap state at $\varepsilon=0$ per dislocation line. Finally, for random $0 \, \leq \, \theta_{\rm d}(\vr_{\rm d}) \!<\! 2 \pi$ the gap is closed and the density of states near $\varepsilon = 0$ is essentially constant. The gap sizes and the occurence of states at energy $\varepsilon=0$ can be heuristically explained by inspecting the phase matching condition for periodic trajectories traveling between the two surfaces at two neighboring dislocation lines at positions $\vr_{\rm d}$ and $\vr_{\rm d}'$. Including the Berry phase $\pi$ for two-dimensional Dirac particles, this phase matching condition reads
\begin{equation}
  \frac{2 \varepsilon}{\hbar v} |\vr_{{\rm d},i} - \vr_{{\rm d},j}| +
  \theta_{\rm d}(\vr_{\rm d}) + \theta_{\rm d}(\vr_{\rm d}') + \pi = 0 \ \ (\mbox{mod}\ 2\pi).
  \label{eq:phase matching}
\end{equation}
Setting $|\vr_{\rm d} - \vr_{\rm d}'| = W/2n$, which is the largest typical distance between neighboring dislocation lines with $2n=6$ dislocations in a slice gives a good estimate of the numerically obtained gap sizes, see Fig.\ \ref{fig:DOS}. The absence of states around $\varepsilon=0$ indicates that pairing of more distant dislocation lines does not occur.

\noindent 

\emph{Conclusion.---} We have investigated the effects of dislocation
line zero modes coupling top- and bottom surfaces of a strong-and-weak
topological insulator slab. Our numerical calculations based on a scattering
approach reveal a rich phenomenology for transport properties and density of
states depending on the phase shifts $\theta_{\rm d}(\vr_{\rm d})$ that
electrons accumulate
along the dislocation lines. For a thin, homogenous slab, a constant
phase shift
$\theta_{\rm d}$ for all dislocation lines can be expected to be a good
approximation. Except for the special cases $\theta_{\rm d}=\pm\pi/2$, this
results in a spectral gap around zero energy and a corresponding strong
suppression of in-plane transport. 
For a thick slab, where dislocation lines are not necessarily straight,
it is conceivable that the phase shifts $\theta_{\rm d}(\vr_{\rm d})$ are
uniformly distributed.
In this case, the in-plane conductivity and the density of states at the nodal energy
are enhanced by the presence of dislocation lines. 

In principle, the dislocation-line-mediated coupling between the top- and bottom surfaces can be described by an effective Hamiltonian involving two Dirac cones coupled by a matrix-valued ``potential''. Such an effective model was considered by Mong {\em et al.} in the context of the transport through a single surface of a weak topological insulator with two (coupled) Dirac cones \cite{mong2012}. The same description can also be applied to the system studied here, although the two Dirac cones now refer to different surfaces. Our analysis shows that the disorder type in such a model depends strongly on the phases accumulated along the dislocation lines: While a mass term is responsible for the opening of a spectral gap (as for $\theta_{\rm d}(\vr_{\rm d}) = \theta_{\rm d}$ constant, $\theta_{\rm d} \neq \pm \pi/2$), a constant scalar potential creates the asymmetry around $\varepsilon = 0$ (which we observe for generic $0 \,\leq \, |\theta_{\rm d}| \! <\! \pi/2$), and zero-average disorder terms lead to the ``flattening'' of the density-of-states singularity at zero energy. Establishing a more rigorous understanding of our results in terms of a Hamiltonian theory would be a formidable task for future work. 

\emph{Acknowledgments.---} We gratefully acknowledge financial support by the Helmholtz Virtual Institute ``New states of matter and their excitations'' and the Alexander von Humboldt foundation.

\bibliographystyle{apsrev}
\bibliography{refs}

\clearpage{}

\onecolumngrid

\section*{\noindent {\large{}Supplemental Material}}

\section{Scattering matrix for scattering off a dislocation line}

\subsection{Polar coordinates\label{sec:polar}}

In the main text we derived the scattering states for a $\pi$ flux line. For a true flux line, the scattering states for a $\pi$ flux and a $-\pi$ flux are identical. Here, we consider the same problem, but include a regularization of the flux line. The regularization will lead to a complete backscattering of the ``$m=0$'' mode, see Eq.\ (\ref{eq:m0}). With regularization, the backscattering phase shift depends on the sign of the flux.

We start with a sharp flux tube for which the vector potential in a polar gauge reads 
\begin{equation}
\mathbf{A}(\mathbf{r})=\frac{hc}{e}\frac{\Phi}{2\pi r}\ve_{\theta},\label{eq:VA}
\end{equation}
where $\Phi$ labels the ``flux'' of the dislocation line in units of the flux quantum (i.e. $\Phi=\pm1/2$ for a $\pm\pi$-flux). With this choice for the vector potential $\mathbf{A}$ the Dirac equation has the form
\begin{equation}
v(\mathbf{p}+e\mathbf{A}/c)\cdot\boldsymbol{\sigma}\psi=-i\hbar v\left(\begin{array}{cc}
0 & D_{-}\\
D_{+} & 0
\end{array}\right)\psi=\varepsilon\psi,\label{eq:DiracSpherical}
\end{equation}
with operators 
\begin{equation}
  D_{\pm}=e^{\pm i\theta}(\partial_{r}\pm\frac{i}{r}\partial_{\theta}\mp\frac{\Phi}{r}).
\end{equation}
Since Eq.\ \eqref{eq:VA} uses a rotationally symmetric gauge the
states can be assumed to be eigenstates of the total angular momentum
$j_{z}=l_{z}+\frac{\hbar}{2}\sigma_{z}$. For $j_{z}=\hbar\left(m-1/2\right)$
they have the form of Eq.\ \eqref{eq:psi_m}, where $m\in\mathbb{Z}$.
The radial part of the wavefunction is then determined by the equations
 \begin{align}
\left(\partial_{r}+\frac{m+\Phi}{r}\right)\phi_{m,-}(r)=ik\phi_{m,+}(r),\\
\left(\partial_{r}-\frac{m+\Phi-1}{r}\right)\phi_{m,+}(r)=ik\phi_{m,-}(r).
\end{align}
where $k=\varepsilon/\hbar v$. 

It is convenient to introduce the kinematic angular momentum
\begin{equation}
j_{z,{\rm kin}}=[\mathbf{r}\times(\mathbf{p}+e\mathbf{A}/c)]_{z}+\frac{\hbar}{2}\sigma_{z}=\hbar\left(m-1/2+\Phi\right)
\end{equation}
The case of interest is $j_{z,{\rm kin}}=0$, {\em i.e.}, $m=1/2-\Phi$.
Here, the radial equations become
\begin{equation}
\left(\partial_{r}+\tfrac{1}{2r}\right)\phi_{\pm}(r)=ik\phi_{\mp}(r),
\end{equation}
where we dropped the index $m$.
These equations are straightforward to solve, and one finds independent
incoming and outgoing radial solutions. 
\begin{equation}
\left(\begin{array}{c}
\phi_{+}(r)\\
\phi_{-}(r)
\end{array}\right)=\alpha_{\mathrm{out}}\frac{1}{\sqrt{r}}e^{ikr}\left(\begin{array}{c}
1\\
1
\end{array}\right)+\alpha_{\mathrm{in}}\frac{1}{\sqrt{r}}e^{-ikr}\left(\begin{array}{c}
1\\
-1
\end{array}\right).\label{eq:spherical_jz0_sol}
\end{equation}
As argued in the main text, the interpretation of the fact that the coefficients $\alpha_{\mathrm{out}}$ and $\alpha_{\mathrm{in}}$ can be chosen independently is that the incoming $j_{z,{\rm kin}}=0$ surface mode is fully transmitted into the outgoing mode in the dislocation line, whereas the incoming dislocation line mode is transmitted in the outgoing surface mode. 

Next, we regularize the $\pi$-flux tube. This requires the breaking of time-reversal symmetry and induces full backscattering of the $j_{z,{\rm kin}}=0$ modes. The simplest way to regularize the flux line using polar coordinates is to take $\Phi$ in Eq.\ \eqref{eq:VA} $r$ dependent. We choose $\Phi(r) = 0$ for $r<\rho$ and $\Phi=\pm1/2$ for $r>\rho$, which corresponds to a situation in which the flux is not located at the origin, but on a circle of radius $\rho$. Obviously, the problem is now well-defined at the origin, and for $r<\rho$ we can take the known solution of the Dirac equation with $\Phi=0$, matching to the solution (\ref{eq:spherical_jz0_sol}) at $r = \rho$. For $\Phi=1/2$ and $m=0$ this procedure yields $\alpha_{\mathrm{out}}=-\alpha_{\mathrm{in}}$. For $\Phi=-1/2$ and $m=1$ we find $\alpha_{\mathrm{out}}=\alpha_{\mathrm{in}}$.

Summarizing: For a regularized flux line the zero-angular-momentum mode is fully backscattered, but with opposite phase factors for a regularization as a positive or as a negative flux. This property will be used in the numerical approach to find the scattering matrix of a dislocation line in Cartesian coordinates, which is outlined in the next Sections.

\subsection{Cartesian coordinates: Scattering off a regularized flux line}

We first consider the scattering problem for a \emph{single} surface Hamiltonian \eqref{eq:H} with a regularized vector potential $\mathbf{A}(\mathbf{r})$. We consider scattering off a string of $2n$ flux lines, all with the same $x$ coordinate $x_{\rm d}$, but with different $y$ coordinates $0<y_{{\rm d},j}<W$, $j=1,2,\ldots,2n$. 

For compatibility with the use of cartesian coordinates, we use a different gauge for the vector potential than in the previous subsection,
\begin{equation}
  \vA(\vr) = \frac{hc}{2 e} \delta(x - x_{\rm d}) \ve_x \alpha(y),
  \label{eq:Ageneral}
\end{equation}
where the function $\alpha(y)$ jumps by one at $y=y_{{\rm d},j}$, see Eq.\ (\ref{eq:Asharp}) of the main text. The vector potential (\ref{eq:Ageneral}) corresponds to the matching condition
\begin{equation}
  \lim_{x \downarrow x_{\rm d}} \psi(x,y) = s(y) \lim_{x \uparrow x_{\rm d}}
  \psi(x,y),
  \label{eq:matching}
\end{equation}
where
\begin{equation}
  s(y) = e^{-i \pi \alpha(y)}.
\end{equation}

\begin{figure}
\begin{centering}
\includegraphics[scale=0.8]{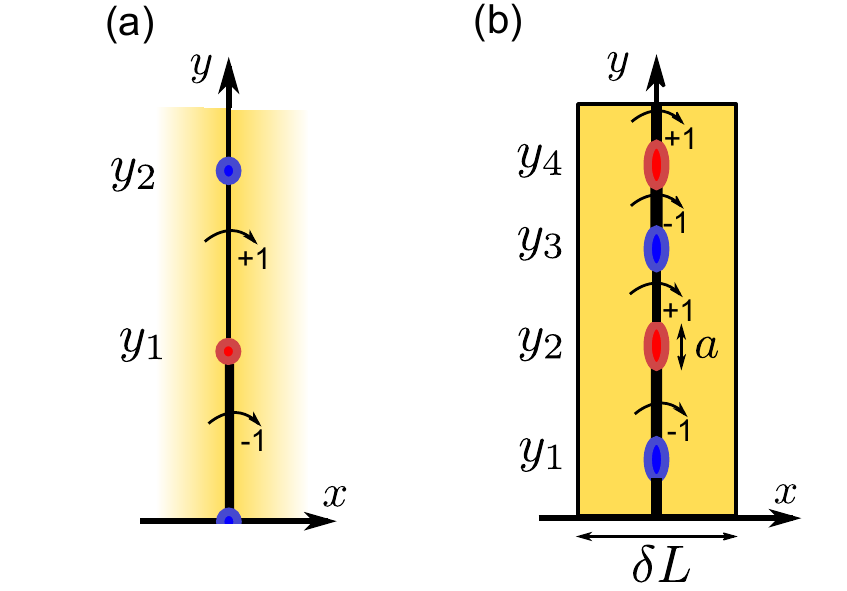}
\par\end{centering}

\protect\caption{{\small{}\label{fig:Setup}Setup of the scattering problem for a string of $2n$ dislocation lines / $\pi$-fluxes located at $x=x_{\rm d}$. The curved arrows denote the phase shift of wavefunctions $s(y)$ that jumps (a) abruptly for sharply defined $\pi$-fluxes or (b) is smeared out over distance $a$ in a regularized setup that also features free propagation in the transport direction of total length $\delta L$.}}
\end{figure}

For sharp flux lines [see Fig. \ref{fig:Setup}(a)], the phase of $s(y)$ changes by $\pi$ at each flux line. Since $s(y)$ is confined to the unit circle in the complex plane, regularization of the flux line corresponds to ``smearing out'' the $\pi$ jumps of the phase factor. There are two possibilities to regularize these jumps: A continuous increase by $\pi$ or a continuous decrease by $\pi$, corresponding to the two signs of the flux in a regularized flux line. In our numerical calculations we smear out each flux line over a distance $a$, see Fig. \ref{fig:Setup}(b), where we make sure that the distance between neighboring flux lines is much larger than $a$. We encode the different regularization possibilities by taking the expression 
\begin{align}
  s_{\tau_{1}...\tau_{2n}}(y)= & -e^{\frac{i\pi}{2}\sum_{j=1}^{2n}\tau_{j}\left[f\left(\frac{y-y_{\mathrm{d},j}}{a}\right)+f\left(\frac{y-y_{\mathrm{d},j}-W}{a}\right)+f\left(\frac{y-y_{\mathrm{d},j}+W}{a}\right)-1\right]},\label{eq:s_s_sp}
\end{align}
where $\tau_{j}=\pm1$ labels the sign of the regularization for the $j$th flux line and the function $f(p)$ is defined as 
\begin{equation}
f(p)=\left\{ \begin{array}{ll}
-1 & \mbox{if \ensuremath{p\le-1}},\\
\tanh(\pi p/\sqrt{1-p^{2}}) & \mbox{if \ensuremath{-1<p<1}},\\
1 & \mbox{if \ensuremath{p\ge1}}.
\end{array}\right.
\end{equation}
The function $f(p)$ is infinitely differentiable at $p=\pm1$. All $2^{2n}$ functions $s_{\tau_{1}...\tau_{2n}}$ have the smeared step as their real part, but different imaginary parts, with positive and negative peaks at each $y_{{\rm d},j}$, thus realizing all $2^{2n}$ possible realizations of the $2n$ flux lines.

The scattering matrix is written in the basis of propagating eigenstates in ideal reference regions immediately to the left and right of the string of flux lines. Following Ref.\ \onlinecite{bardarson2007}, we take the Hamiltonian for these reference regions as $H_{\rm ref} = v p_x \sigma_x$. The omission of the term $v p_y \sigma_y$ is inconsequential, since the reference regions are used for reference purposes only and their length is sent to zero at the end of the calculation. In each reference region we expand the wavefunction $\psi(\vr)$ in basis states $|q_{n},\pm\rangle$, where the sign $\pm$ refers to right-/leftmoving states, and $q_{n}=2\pi n/W$ labels the transverse momenta for periodic boundary conditions. The corresponding wavefunctions are
\begin{equation}
  \phi_{n,\pm}(x,y) = \langle x,y |q_{n},\pm\rangle = 
  \frac{1}{\sqrt{2W}}e^{\pm i k x + iq_{n}y}
  \left(\begin{array}{c} 1\\ \pm 1
\end{array}\right).\label{eq:leadModes}
\end{equation}

To obtain the scattering matrix of the string of (regularized) dislocation lines, we solve for scattering states of the form
\begin{equation}
  \psi(x,y) 
  =\begin{cases}
  \sum_{n} \left[ a_{{\rm L},n}^{\rm in} \phi_{n,+}(x,y) +
  a_{{\rm L},n}^{\rm out} \phi_{n,-}(x,y) \right] & \mathrm{for}\;\;x<x_{\rm d},\\
  \sum_{n} \left[ a_{{\rm R},n}^{\rm in}  \phi_{n,-}(x,y)
 + a_{{\rm R},n}^{\rm out} \phi_{n,+}(x,y) \right] & \mathrm{for}\;\;x>x_{\rm d},\\
\end{cases}
  \label{eq:psiplus}
\end{equation}
and obtain the scattering matrix from the linear relation
\begin{equation}
  \left( \begin{array}{c} a_{L}^{\rm out} \\ a_{R}^{\rm out} \end{array} \right)
  = S \left( \begin{array}{c} a_{L}^{\rm in} \\ a_{R}^{\rm in} \end{array} \right), \ \
  S = \left( \begin{array}{cc} r & t' \\ t' & r \end{array} \right).
\end{equation}
The amplitudes $r_{nm}$, $r'_{nm}$, $t_{nm}$, and $t'_{nm}$ of the reflection and transmission blocks of the scattering matrix can be calculated from the matching condition (\ref{eq:matching}). Since the matching condition does not relate to the pseudospin and only affects the phase of the wavefunction, we directly conclude that there is no reflection caused by the dislocation line, 
\begin{equation}
  r_{nm}=r'_{nm}=0.
\end{equation}
To obtain $t_{nm}$, we substitute Eq.\ (\ref{eq:psiplus}) into Eq.\ (\ref{eq:matching}) and perform a Fourier transform to $y$,
\begin{eqnarray}
  t_{nm} &=& t_{mn}'^* = \frac{1}{W}\int_{0}^{W}dys(y)e^{i(q_{m}-q_{n})y} .
\end{eqnarray}
We have suppressed the dependence of the transmission matrices $t$ and $t'$ and the function $s(y)$ on the flux regularization parameters $\tau_j$ to keep our notation simple.

Up to this point the number of transverse modes has been infinite. For a practical implementation, we need to employ a mode cutoff $N$ such that $n=-N+1,...,0,...,N-1$. Naive truncation of the transmission matrices $t$ and $t'$, however, leads to a non-unitary scattering matrix. To circumvent this problem, we add segments of a finite width $\delta L/2$ to the left and to the right of the impurity lines, as shown in Fig.\ \ref{fig:Setup}(b). The scattering matrix for such slices are known. For zero energy the reflection and transmission amplitudes $\rho_{nm}$ and $\tau_{nm}$ read \cite{tworzydlo2006} 
\begin{align}
  \rho_{nm} & =-\rho'_{nm} = -\tanh(q_{n}\delta L/2)\delta_{nm}, \nonumber \\
  \tau_{nm} & =\tau^{\prime}{}_{nm} =\frac{1}{\cosh(q_{n}\delta L/2)}\delta_{nm}.
\end{align}
Since modes with high momenta $q_{n}\gg 2/\delta L$ are blocked from propagation, we should be allowed to safely truncate the scattering matrix of the dislocation line string with the two dislocation-line free sigments of length $\delta L/2$ on each side, once $N$ is much larger than $W/\delta L$. Thus, we consider the concatenated scattering matrix of a single surface in a geometry of Fig. \ref{fig:Setup}(b) with regularized fluxes which reads 
\begin{eqnarray}
S_{\mathrm{reg}}^{\tau_{1}...\tau_{2n}}	=	\left(\begin{array}{cc}
\rho & \tau^{\prime}\\
\tau & \rho^{\prime}
\end{array}\right)\otimes\left(\begin{array}{cc}
0 & t^{\prime\tau_{1}...\tau_{2n}}\\
t^{\tau_{1}...\tau_{2n}} & 0
\end{array}\right)\otimes\left(\begin{array}{cc}
\rho & \tau^{\prime}\\
\tau & \rho^{\prime}
\end{array}\right)
	=	\left(\begin{array}{cc}
\rho+\tau't'\frac{1}{1-\rho t\rho't'}\rho t\tau & \tau't'\frac{1}{1-\rho t\rho't'}\tau'\\
\tau\frac{1}{1-t\rho't'\rho}t\tau & \rho'+\tau\frac{1}{1-t\rho't'\rho}t\rho't'\tau'
\end{array}\right)\label{eqn:Sreg}
\end{eqnarray}
where $\otimes$ denotes concatenation of scattering matrices and we restored the regularization indices $\tau_j$. For sufficiently large $N$ the matrix $S_{\mathrm{reg}}^{\tau_{1}...\tau_{2n}}$ is unitary to within our numerical accuracy.

\noindent 

\subsection{Structure of regularized scattering matrices}

The calculation of Sec.\ \ref{sec:polar} showed, that in the limit $\delta L\gg a$ the choice of the regularization of the dislocation lines affects precisely one mode. This can also be verified numerically for plane-wave scattering states. 

For the numerical analysis, we take $a \ll \delta L \ll d$, with $d = \min_{j}|y_{{\rm d},j} - y_{{\rm d},j-1}|$ the minimum distance between neighboring flux lines, and choose the mode cut-off $N \gg W/\delta L$. We then find that the {\em difference} 
\begin{equation}
  \Delta S_{\rm reg}^{(j)} \equiv
  S_{{\rm reg}}^{\tau_1,\ldots,\tau_{j-1},+,\tau_{j+1},\ldots,\tau_{2n}}-S_{{\rm reg}}^{\tau_1,\ldots,\tau_{j-1},-,\tau_{j+1},\ldots,\tau_{2n}}
\end{equation}
is (i) independent of the regularization parameters $\tau_1,\ldots,\tau_{j-1},\tau_{j+1},\ldots,\tau_{2n}$, (ii) of unit rank, and (iii) with norm two.  Hence, there exists a normalized vector $v_{j}$ such that
\begin{equation}
   \Delta S_{\rm reg}^{(j)}
  = 2 v_j v_j^{\dagger},
  \label{eq:Sregtau}
\end{equation}
independent of $\tau_1,\ldots,\tau_{j-1},\tau_{j+1},\ldots,\tau_{2n}$. (Since at zero energy the problem at hand has chiral symmetry, $\sigma_z H \sigma_z = -H$, the scattering matrix is hermitian and the single non-vanishing eigenvalue of $\Delta S_{\rm reg}^{(j)}$ has to be $\pm2$. For our choice of the function $s(y)$, we find the positive sign realized.)
The interpretation of this result is that difference $\Delta S_{\rm reg}^{(j)}$
relates to the choice of the regularization of the $j$th flux line only. Since the change of the regularization of the $j$th flux line changes the sign of the scattering amplitude of the zero angular momentum mode (defined with respect to the $j$th flux line) and leaving all other scattering amplitudes unchanged, the difference  $\Delta S_{\rm reg}^{(j)}$
precisely describes that contribution to the total scattering matrix $S_{\rm reg}$ that originates from scattering of the zero angular momentum mode for the $j$th flux line off that same flux line. As long as the separation between flux lines is much larger than the slice width $\delta L$, contributions from different flux lines do not interfere, which is why  $\Delta S_{\rm reg}^{(j)}$ and, hence, the vector $v_{j}$ is independent of the regularization parameters $\tau_1,\ldots,\tau_{j-1},\tau_{j+1},\ldots,\tau_{2n}$ of the remaining $2n-1$ flux lines.

Repeating this procedure for all $2n$ flux lines, we find that we can write
\begin{equation}
S_{\mathrm{reg}}^{\tau_{1}...\tau_{2n}}=S_{0}+\sum_{j=1}^{2n}\tau_{j}\, v_{j}v_{j}^{\dagger}\label{eq:SregSum}
\end{equation}
with 
\begin{equation}
S_{0}= \frac{1}{2} (S_{{\rm reg}}^{+...+}+S_{{\rm reg}}^{-...-})\label{eq:S0}
\end{equation}
the part of the scattering matrix that describes transport not affected by the choice of regularization of any dislocation line. It has rank $2 N - 1 - 2n$. In keeping with the above interpretation, the matrix $S_{0}$ describes scattering from the non-zero-angular momentum modes, whereas the term $v_{j} v_{j}^{\dagger}$ describes the contribution to $S_{\rm reg}$ from the zero-angular momentum mode at the $j$th flux line.

\begin{figure}[t]
\centering{}
\includegraphics[scale=0.8]{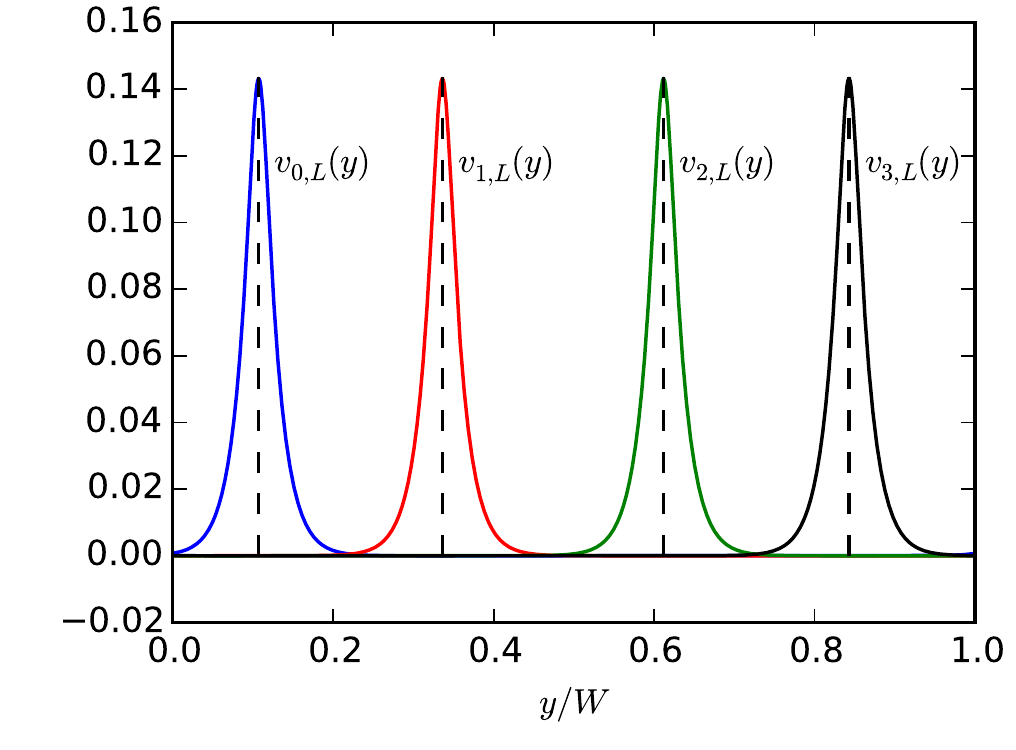}
\protect\caption{{\small{}\label{fig:realSpace}Fourier transform of the first $2N+1$
entries $v_{j,L}$ of the vectors $v_{j}$, $j=1,2,3,4$ ($2n=4$)
that describes the real space wavefunction of the eigenmodes scattering
at the dislocation lines at $y=y_{\mathrm{d},j}$ at the left lead
at $x=-\delta L/2$. The parameters are $\delta L=0.03W$, $a=0.3\delta L$,
$N=400$.}}
\end{figure}

It is instructive (though inessential for future calculations) to
look at the Fourier spectrum of the vectors $v_{j}$ whose first (second)
$2N+1$ entries encode the real space structure of the eigenmodes
scattering at the dislocation lines at position $x=\mp\delta L/2$
at the left (right) lead. The Fourier transforms are real and depicted
in Fig. \ref{fig:realSpace} where dashed lines indicate the positions
of the dislocation lines. 

\noindent 

\subsection{Time-reversal symmetry}

The scattering matrices $S_{\mathrm{reg}}^{\tau_{1}...\tau_{2n}}$ are not time-reversal symmetric because of the presence of the smeared flux line. However, the matrix $S_{0}$ is time-reversal symmetric. We here summarize how time reversal symmetry is implemented in the present problem. 

The time-reversal operator is ${\cal T}=i\sigma_{y}K$, with $K$ complex conjugation. It satisfies ${\cal T}^{2}=-1$. 
Time reversal symmetry applied to basis states gives 
\begin{equation}
{\cal T}\left|q_{n},\pm\right\rangle =\pm\left|-q_{n},\mp\right\rangle.
\end{equation}
If the Hamiltonian is time-reversal symmetric, the scattering matrix
satisfies $\langle f|\hat{S}|i\rangle =\langle {\cal T}i|\hat{S}|{\cal T}f\rangle$. From that, one finds the conditions
\begin{align*}
t_{nm} & =t_{-m,-n}^{\prime},\ \
r_{nm} =-r_{-m,-n},\ \
r_{nm}^{\prime}  =-r_{-m,-n}^{\prime}
\end{align*}
for the reflection and transmission amplitudes. One rewrite these equations using the matrix 
\begin{equation}
U=\left(\begin{array}{cccc}
0 & \dots & 0 & 1\\
0 & \dots & 1 & 0\\
\vdots &  &  & \vdots\\
1 & 0 & \dots & 0
\end{array}\right)
  \label{eq:Udef}
\end{equation}
with $U^{2}=1$, which switches from positive to negative momenta. Then:
\begin{align}
t & =Ut^{\prime{\rm T}}U, \ \
r  =-Ur^{{\rm T}}U,\nonumber \ \
r^{\prime}  =-Ur^{\prime{\rm T}}U,
\end{align}
or, equivalently, 
\begin{equation}
S=-\left(\begin{array}{cc}
U & 0\\
0 & -U
\end{array}\right)S^{{\rm T}}\left(\begin{array}{cc}
U & 0\\
0 & -U
\end{array}\right).
\end{equation}

As remarked above, the scattering matrices $S_{{\rm reg}}^{\sigma\sigma'}$ are not time-reversal symmetric because of the presence of the smeared flux tube. However, the matrix $S_{0}$ is time-reversal symmetric. Similarly, the difference $\Delta S_{\rm reg}^{(j)}$ acquires a minus sign under time reversal. This property can be used to remove the over-all phase factor ambiguity of the vectors $v_{j}$ [the phase was not specified in the definition (\ref{eq:SregSum})], up to a sign ambiguity,
\begin{equation}
  v_{j} = \left(\begin{array}{cc}
U & 0\\
0 & -U
\end{array}\right) v_{j}^*.
  \label{eq:VU}
\end{equation}

\subsection{Scattering scattering matrix $\tilde{S}$ for a thin slice}

The key element of our calculation of the scattering scattering matrix $\tilde{S}$ for a thin slice is that a regularization of the $\pi$-flux lines does not affect the way angular modes with nonzero (kinetic) angular momentum are scattered off the flux line, see Eq.\ (\ref{eq:phi}), but that regularization does lead to full backscattering of the zero angular momentum modes. In the real weak-and-strong topological insulator, it is these latter modes that are fully transmitted from the surface into the dislocation line and vice versa. 

As we have discussed above, the backscattering phase shift for the zero angular momentum mode depends on whether one chooses to regularize the $\pi$-flux line with a magnetic field in the positive $z$ direction, or with a magnetic field in the negative $z$ direction. By calculating the scattering matrices with different regularizations of the $\pi$-flux lines, we can separate the contributions from angular modes with nonzero angular momentum, which are independent of the regularization and which do not transmit into the dislocation line from modes with zero angular momentum, which are dependent on the regularization, and which are fully transmitted into the dislocation line. The decomposition (\ref{eq:SregSum}) allows us to uniquely separate these different contributions to the scattering matrix.

These arguments can be used to construct a larger scattering matrix for a surface and the $2n$ helical modes corresponding to the $2n$ dislocation lines piercing the surface at $x=x_{\rm d}$. We denote the amplitudes for the surface-state modes in the top surface by $a_{{\rm L,R},+}^{\rm in,out}$, and the amplitudes for surface-state modes in the bottom surface by $a_{{\rm L,R},-}^{\rm in,out}$. We denote the amplitudes of the upward and downward traveling helical modes in the $j$th dislocation line at the top surface by $b_{j,+}^{\rm in}$ and $b_{j,+}^{\rm out}$, respectively, and we denote the amplitudes of the upward and downward traveling helical modes in the $j$th dislocation line at the bottom surface by $b_{j,-}^{\rm out}$ and $b_{j,-}^{\rm in}$, respectively. Following the above arguments, the scattering matrix $\tilde S_+$ relating the surface states in the top surface and the helical modes along the dislocation lines at the top surface then reads
\begin{equation}
\left(\begin{array}{c}
a_{{\rm L,R}+}^{\mathrm{out}}\\
b_{1,+}^{\mathrm{out}}\\
\vdots \\
b_{2n,+}^{\mathrm{out}}
\end{array}\right)
  =
  \underset{\equiv\tilde{S}_{+}}{\underbrace{\left(\begin{array}{cccc}
  S_{0} & v_{1} & \ldots & v_{2n}\\
  e^{i \phi_1} v_{1}^{\dagger} & 0 & \ldots & 0\\
  \vdots & \vdots & & \vdots \\
  e^{i \phi_{2n}} v_{2n}^{\dagger} & 0 & \ldots & 0
  \end{array}\right)}}
  \left( \begin{array}{c} 
  a_{{\rm L,R}+}^{\mathrm{in}}\\
  b_{1,+}^{\mathrm{in}} \\ \vdots \\
  b_{2n,+}^{\mathrm{in}}
\end{array} \right).
  \label{eq:Stplus}
\end{equation}
A similar expression can be found for the scattering matrix $\tilde S_{-}$ relating the surface states in the top surface and the helical modes along the dislocation lines at the bottom surface. The phases $\phi_{j}$ can not be determined using the above arguments. Instead, we can determine these phases from the condition of time-reversal symmetry. With the convention 
\begin{equation}
  {\cal T}|{\rm in}\rangle =-|{\rm out}\rangle \quad\mathrm{and}\quad{\cal T}|{\rm out}\rangle ={\rm |{\rm in}\rangle }
\label{eq:TRS_1d}
\end{equation}
for the time-reversal symmetry operation $\mathcal{T}$ (with $\mathcal{T}^{2}=-1$) on the in and outgoing dislocation line states $|{\rm in}\rangle $ and $|{\rm out}\rangle $, we find that $\tilde S_+$ must obey the condition
\begin{equation}
  \tilde S_+ = -
  \left( \begin{array}{ccc} U & 0 & 0 \\ 0 & -U & 0 \\ 0 & 0 & -1_{2n} 
  \end{array} \right)
  \tilde S_+^{\rm T}
  \left( \begin{array}{ccc} U & 0 & 0 \\ 0 & -U & 0 \\ 0 & 0 & -1_{2n} 
  \end{array} \right),
\end{equation}
where the matrix $U$ was defined in Eq.\ (\ref{eq:Udef}) and where $1_{2n}$ is the $2n \times 2n$ unit matrix. Comparison with Eq.\ (\ref{eq:Stplus}) and using Eq.\ (\ref{eq:VU}) then gives that $\phi_{j}=0$.

\noindent 
\begin{figure}
\begin{centering}
\includegraphics[scale=0.8]{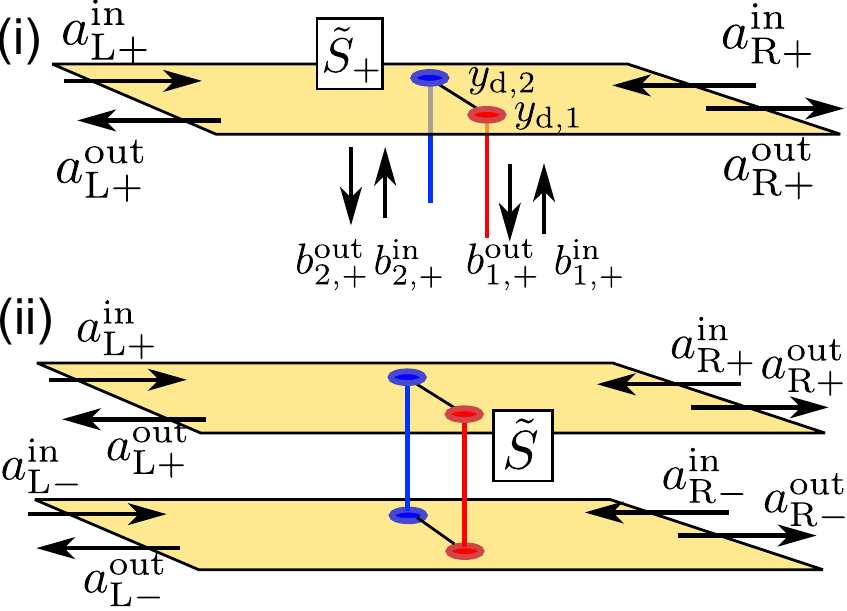}
\par\end{centering}

\protect\caption{{\small{}\label{fig:LargeScatteringMat}(i) Scattering setup for two
dislocation lines impinging on a top surface. The dislocation lines
are explicitly treated as terminals in the unitary scattering relation
Eq. \eqref{eq:Stplus} involving the scattering matrix $\tilde{S}_{+}$.
(ii) Enlarged scattering matrix $\tilde{S}$ describing scattering
between top and bottom surface via a pair of dislocation lines. }}
\end{figure}

In the next step we connect the scattering matrices $\tilde{S}_{+}$ and $\tilde{S}_{-}$ to obtain the scattering matrix $\tilde{S}$ describing the scattering off a string of dislocation lines of surface states in {\em both} surfaces of the weak-and-strong topological insulator. The procedure is shown schematically in Fig. \ref{fig:LargeScatteringMat}(ii). In order to connect the two layers we need the additional requirement
\begin{eqnarray}
b_{j,+}^{{\rm in}} & = & e^{i\theta_{\mathrm{d},j}}b{}_{j,-}^{\mathrm{out}},\\
b_{j,+}^{{\rm out}} & = & -e^{-i\theta_{\mathrm{d},j}}b{}_{j,-}^{\mathrm{in}},
\end{eqnarray}
which relates the helical-state amplitudes at the top and bottom layers. The phase $\theta_{\mathrm{d},j}$ describes the phase accumulated during the propagation along the dislocation line. The minus sign ensures that the time-reversal convention Eq.\ \eqref{eq:TRS_1d} applies to each layer separately.
Eliminating the amplitudes for the dislocation line and using the same sign choice for $v_j$ for top and bottom layer, \emph{c.f.} Eq. \eqref{eq:VU}, one arrives at the scattering matrix
\begin{equation}
\left(\begin{array}{c}
a_{\mathrm{L}+}^{\mathrm{out}}\\
a_{{\rm R}+}^{\mathrm{out}}\\
a_{\mathrm{L}-}^{\mathrm{out}}\\
a_{{\rm R}-}^{\mathrm{out}}
\end{array}\right)=
  \tilde S
\left(\begin{array}{c}
a_{\mathrm{L}+}^{\mathrm{in}}\\
a_{{\rm R}+}^{\mathrm{in}}\\
a_{\mathrm{L}-}^{\mathrm{in}}\\
a_{{\rm R}-}^{\mathrm{in}}
\end{array}\right),\ \
  \tilde S = \left(\begin{array}{cc}
S_{0} & \sum_{j=1}^{2n}e^{i\theta_{{\rm d},j}}v_{j}v_{j}^{\dagger}\\
-\sum_{j=1}^{2n}e^{i\theta_{{\rm d},j}}v_{j}v_{j}^{\dagger} & S_{0}
\end{array}\right).
  \label{eq:St}
\end{equation}
One easily verifies that this scattering matrix is time-reversal symmetric.

\section{Transport properties}

\begin{figure}
\begin{centering}
\includegraphics{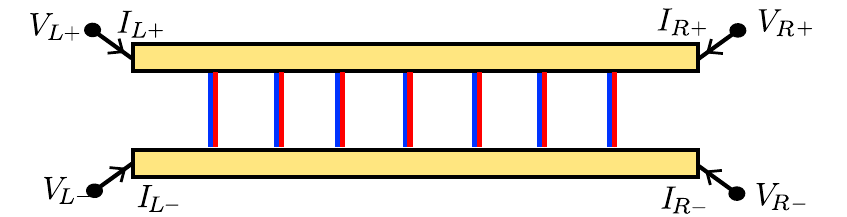} 
\par\end{centering}
\protect\caption{\label{fig:top-side-view2}
Schematic picture of a sideview of the topological insulator slab, together with the definitions of the potentials and currents at the four contacts.
}
\end{figure}

For a description of transport properties, four ideal contacts are added to the top and bottom surfaces for $x<0$ and $x>L$. Following Ref.\ \onlinecite{tworzydlo2006} these are described by the Hamiltonian $H_{\rm ref} = v p_x \sigma_x$. We do not place dislocation lines in the contact regions, so that the scattering states in the contacts remain unaffected by the presence of the dislocation lines. Voltages  $V_{{\rm L}\sigma}$ and $V_{{\rm R}\sigma}$ and currents $I_{{\rm L}\sigma}$ and $I_{{\rm R}\sigma}$  at the four contacts are defined as shown in Fig.\ \ref{fig:top-side-view2}.

The different transport properties require different configurations
for the voltages at the four contacts. 
The in-plane conductance $G_{\parallel}=I_{\parallel}/V_{\parallel}$ is obtained by setting $V_{{\rm L}+}=V_{{\rm L}-}=V_{\parallel}$ and $V_{{\rm R}+}=V_{{\rm R}-}=0$ and measuring the current $I_{\parallel}=I_{{\rm R}+}+I_{{\rm R}-}=I_{{\rm L}+}+I_{{\rm L}-}$.
The cross conductance $G_{\perp}=I_{\perp}/V_{\perp}$ between the bottom and top surface is defined by setting $V_{{\rm L}+}=V_{{\rm R}+}=V_{\perp}$, $V_{{\rm L}-}=V_{{\rm R}-}=0$, and measuring the total current $I_{\perp}=I_{{\rm L}-}+I_{{\rm R}-}=-(I_{{\rm L}+}+I_{{\rm R}+})$. 
Expressions for $G_{\parallel}$ and $G_{\perp}$ in terms of the scattering matrix and the results of numerical calculations of  $G_{\parallel}$ and $G_{\perp}$ are given in the main text.

In addition to $G_{\parallel}$ and $G_{\perp}$ we also considered the drag resistance $R_{{\rm d}}=V_{{\rm d}}/I_{{\rm d}}$, which is the ratio of an open-circuit voltage at one surface induced by an applied current at the other surface. It is based on the configuration $I_{{\rm L}+}=-I_{{\rm R}+}=I_{{\rm d}}$, $I_{{\rm L}-}=I_{{\rm R}-}=0$, $V_{{\rm R}-}=0$, and $V_{{\rm L}-}=V_{{\rm d}}$. For the calculation of the drag resistance $R_{{\rm d}}$ we start from the conductance matrix connecting a generic configuration of voltages and currents. Using current conservation and the reference $V_{R-}\equiv0$, we have
\begin{equation}
\left(\begin{array}{c}
I_{\mathrm{L}+}\\
I_{\mathrm{L}-}\\
I_{{\rm R}+}
\end{array}\right)=\left(\begin{array}{ccc}
G_{\mathrm{L}+,\mathrm{L}-}+G_{\mathrm{L}+,{\rm R}+}+G_{\mathrm{L}+,{\rm R}-} & -G_{\mathrm{L}+,\mathrm{L}-} & -G_{\mathrm{L}+,{\rm R}+}\\
-G_{\mathrm{L}-,\mathrm{L}+} & G_{\mathrm{L}-,\mathrm{L}+}+G_{\mathrm{L}-,{\rm R}+}+G_{\mathrm{L}-,{\rm R}-} & -G_{\mathrm{L}-,{\rm R}+}\\
-G_{{\rm R}+,\mathrm{L}+} & -G_{{\rm R}+,\mathrm{L}-} & G_{{\rm R}+,\mathrm{L}+}+G_{{\rm R}+,\mathrm{L}-}+G_{{\rm R}+,{\rm R}-}
\end{array}\right)\left(\begin{array}{c}
V_{\mathrm{L}+}\\
V_{\mathrm{L}-}\\
V_{{\rm R}+}
\end{array}\right)\label{eq:G-mat}
\end{equation}
where, in terms of the scattering matrix defined in Eq. \eqref{eq:Ssigmasigmap},
\begin{eqnarray}
G_{\mathrm{L}\sigma,\mathrm{L}\sigma^{\prime}} & = & \frac{e^{2}}{h}\mathrm{tr}\, r_{\sigma\sigma^{\prime}}r_{\sigma\sigma^{\prime}}^{\dagger},\\
G_{\mathrm{L}\sigma,{\rm R}\sigma^{\prime}} & = & \frac{e^{2}}{h}\mathrm{tr}\, t_{\sigma\sigma^{\prime}}^{\prime}t_{\sigma\sigma^{\prime}}^{\prime\dagger},\\
G_{{\rm R}\sigma,\mathrm{L}\sigma^{\prime}} & = & \frac{e^{2}}{h}\mathrm{tr}\, t_{\sigma\sigma^{\prime}}t_{\sigma\sigma^{\prime}}^{\dagger},\\
G_{{\rm R}\sigma,{\rm R}\sigma^{\prime}} & = & \frac{e^{2}}{h}\mathrm{tr}\, r_{\sigma\sigma^{\prime}}^{\prime}r_{\sigma\sigma^{\prime}}^{\prime\dagger}.
\end{eqnarray}
Inverting Eq. \eqref{eq:G-mat}, the resistance matrix is obtained as
\begin{equation}
\left(\begin{array}{c}
V_{\mathrm{L}+}\\
V_{\mathrm{L}-}\\
V_{{\rm R}+}
\end{array}\right)=\left(\begin{array}{ccc}
R_{\mathrm{L}+,\mathrm{L}+} & R_{\mathrm{L}+,\mathrm{L}-} & R_{\mathrm{L}+,{\rm R}+}\\
R_{\mathrm{L}-,\mathrm{L}+} & R_{\mathrm{L}-,\mathrm{L}-} & R_{\mathrm{L}-,{\rm R}+}\\
R_{{\rm R}+,\mathrm{L}+} & R_{{\rm R}+,\mathrm{L}-} & R_{{\rm R}+,{\rm R}+}
\end{array}\right)\left(\begin{array}{c}
I_{\mathrm{L}+}\\
I_{\mathrm{L}-}\\
I_{{\rm R}+}
\end{array}\right)\label{eq:R_mat}
\end{equation}
and the configuration specified for the drag resistance ($I_{{\rm L}+}=-I_{{\rm R}+}=I_{{\rm d}}$,
$I_{{\rm L}-}=I_{{\rm R}-}=0$, $V_{{\rm R}-}=0$, and $V_{{\rm L}-}=V_{{\rm d}}$)
can be applied. Solving for $R_{{\rm d}}=V_{{\rm d}}/I_{{\rm d}}$
yields
\begin{equation}
R_{\mathrm{d}}=V_{\mathrm{d}}/I_{\mathrm{d}}=R_{\mathrm{L}-,\mathrm{L}+}-R_{\mathrm{L}-,{\rm R}+}.
\end{equation}

\begin{figure}
\begin{centering}
\includegraphics{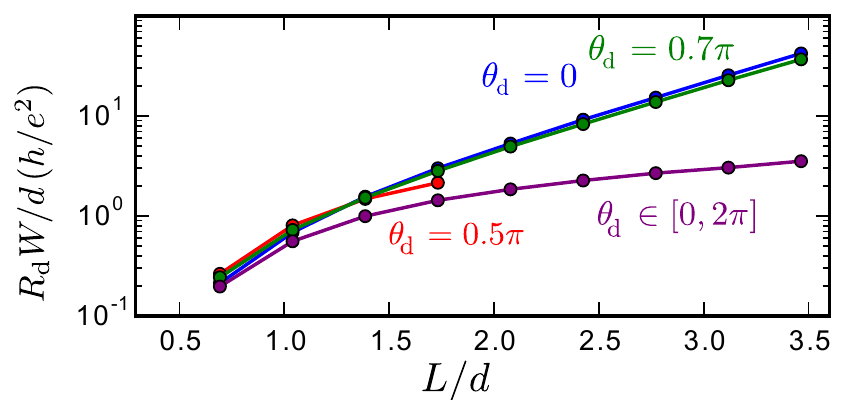}
\par\end{centering}

\protect\caption{\label{fig:results_transport2}
Zero-energy drag resistance $R_{{\rm d}}$ for a slab of weak-and-strong topological insulator with a concentration $1/d^{2}$ of randomly placed dislocation lines. The different curves refer to different choices for the phases  $\theta_{{\rm d}}(\vr_{\rm d})$, as shown in the figure. 
Data points denote an average over 500 disorder realizations, statistical
error bars are typically smaller than the markers.}
\end{figure}

Figure \ref{fig:results_transport2} shows the ensemble averaged drag resistance, multiplied by $d/W$ to remove a trivial dependence on the sample width. Analogous to the in-plane and cross conductances discussed in the main text, there is a strong dependence on the choice of the phases $\theta_{{\rm d}}(\vr_{\rm d})$ accumulated along the dislocation lines.

\section{Parameters for simulation}

For the numerical simulation, we chose $W/L=5$ for the aspect ratio
of the slab, we verified that this is large enough that $G_{\parallel,\perp}$ and $R_{\mathrm{d}}$ were proportional and
inversely proportional to $W$, respectively. We divided the slab
in transport direction in 10 slices with $2n=6$ dislocation lines
each, so that $d/W=0.058$. For each slice of width $\delta L$, 
the scattering matrix is calculated with mode cutoff $N=600$. For 
the concatenation of scattering matrices of different slices we 
imposed a smaller cut-off $N_{\mathrm{eff}}=65$ for the number of
modes. We verified that $N$ and $N_{\rm eff}$ are large enough that 
the results do not depend on these numbers.

\section{Density of states}

We discuss how the density of states can be calculated for a periodic array of 
segments of length $L$. Equivalently, one may apply ``twisted''
boundary conditions in the $x$ direction, in which electrons
pick up an additional phase $e^{i \kappa_{x} L}$, $\kappa_x$ being
the crystal momentum, while passing across the ``boundary''.

The procedure is illustrated
in Fig. \ref{fig:Calculation-of-DOS}.
We start from the scattering matrix of the open slice at energy $\varepsilon$,
which we calculate as described previously,
\[
\left(\begin{array}{c}
L_{\rm out}\\
R_{\rm out}
\end{array}\right)=\underset{S\left( \varepsilon \right)}{\underbrace{\left(\begin{array}{cc}
r & t^{\prime}\\
t & r^{\prime}
\end{array}\right)}}\left(\begin{array}{c}
L_{\rm in}\\
R_{\rm in}
\end{array}\right).
\]%
The matching conditions on the in- and outgoing states for a Bloch state with crystal momentum $\kappa_{x}$ read
\[
\left(\begin{array}{c}
R_{\rm out}\\
L_{\rm out}
\end{array}\right)=\left(\begin{array}{c}
e^{i \kappa_x L}L_{\rm in}\\
e^{-i \kappa_x L}R_{\rm in}
\end{array}\right)
\]
which has a non-trivial solution (indicating an eigenstate of the
closed system at energy $\varepsilon$) if and only if the matrix 
\[
S_{\kappa_x}\left(\varepsilon \right)\equiv\left(\begin{array}{cc}
0 & e^{-i \kappa_x L}\\
e^{i\kappa_x L} & 0
\end{array}\right)S\left(\varepsilon \right)
\]%
has a unit eigenvalue. In practice, since the matrix $S_{\kappa_x}\left(\varepsilon\right)$
is unitary, we track the eigenvalue phases with varying $\varepsilon$ and identify
states at energies where a phase crosses zero. The phase
$\kappa_x L$ controls the boundary condition in $x$ direction and averaging over $\kappa_x L \in\left[0,2\pi\right]$ thus reduces finite-size
effects (we used 80 equally spaced values for $\kappa_x L$ from the interval
$\left[0,2\pi\right]$). A similar procedure could be applied with
a phase controlling the boundary conditions in transversal direction.

\begin{center}
\begin{figure}
\begin{centering}
\includegraphics{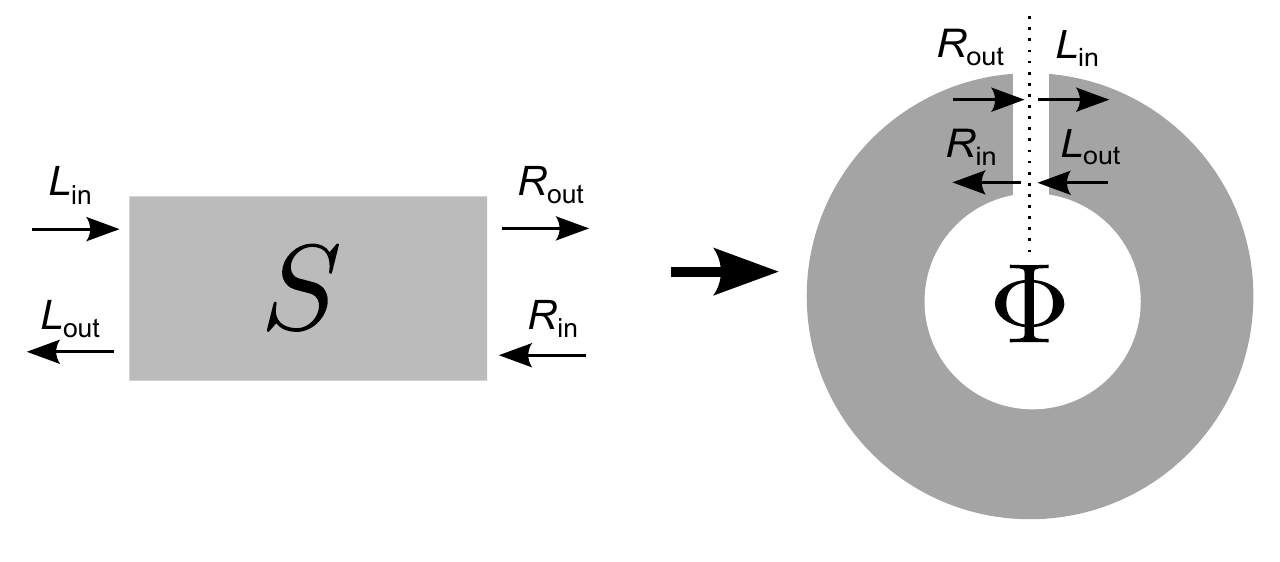}
\par\end{centering}

\protect\caption{\label{fig:Calculation-of-DOS}
Calculation of density of states of a periodic array
(right) from scattering matrix $S(\varepsilon)$ of an open system (left). The
incoming and outcoming states should match, up to a 
phase factor $\Phi=\kappa_x L$ from the crystal momentum. }
\end{figure}

\par\end{center}

\section{Additional choices for the phases $\theta_{\rm d}(\vr_{\rm d})$}

\begin{figure}
\begin{centering}
\includegraphics{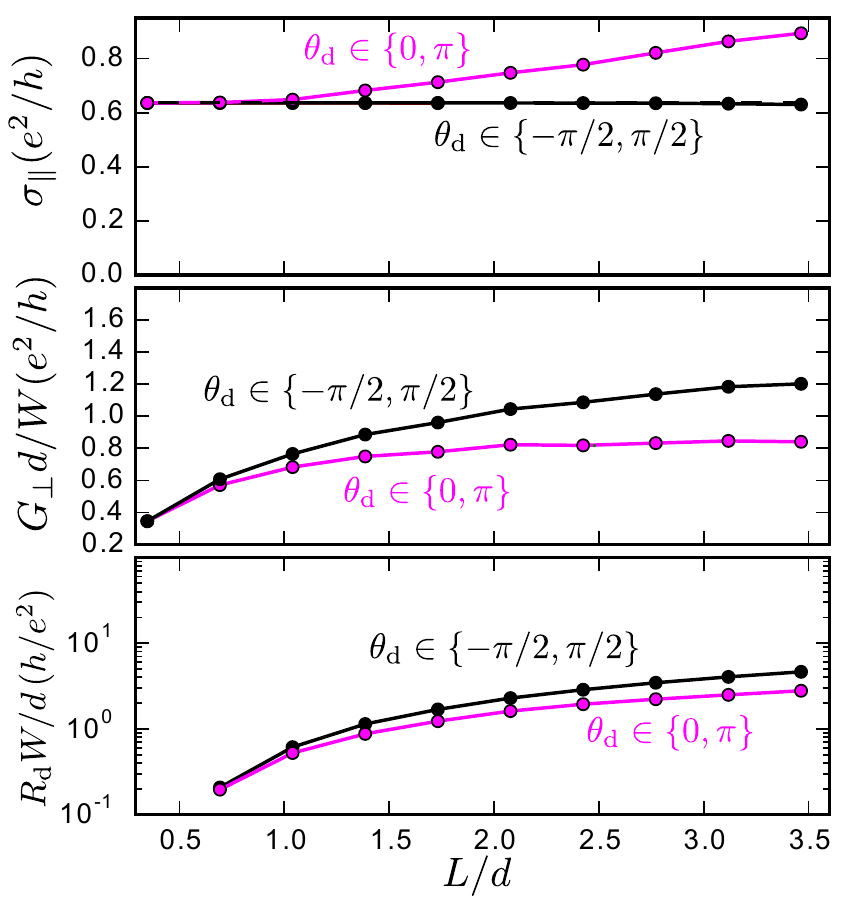}
\par\end{centering}

\protect\caption{\label{fig:results_transport3}
In-plane conductivity $\sigma_{\parallel}$
(a), cross conductance $G_{\perp}$ (b), and drag resistance $R_{{\rm d}}$ (c) for a slab of weak-and-strong topological insulator with a concentration $1/d^{2}$ of randomly placed dislocation lines. The different curves refer to the phases  $\theta_{{\rm d}}(\vr_{\rm d})$ randomly chosen from $\{0,\pi\}$ or from $\{-\pi/2,\pi/2\}$, as indicated in the figure. 
Data points denote an average over 500 disorder realizations, statistical
error bars are typically smaller than the markers.}
\end{figure}

\begin{figure}
\noindent \begin{centering}
\includegraphics{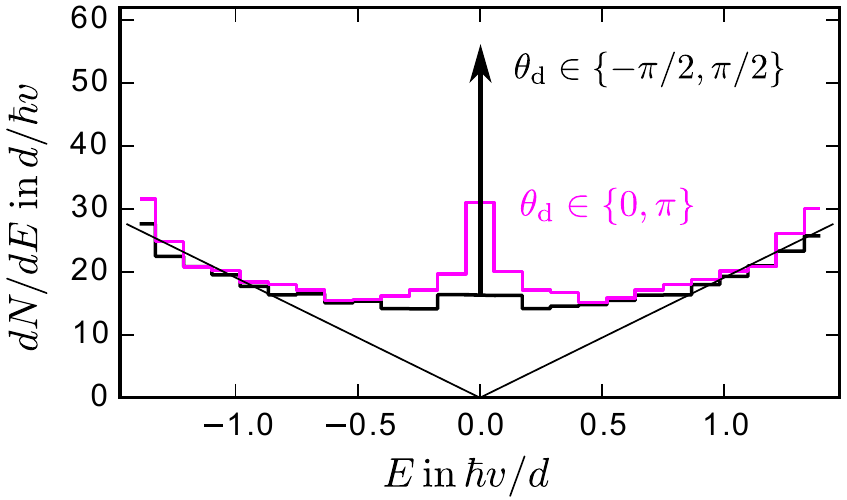}
\par\end{centering}

\protect\caption{\label{fig:dos2}Density of states $dN/dE$ of a sample with dislocation line density $1/d^{2}$. The two curves represent are for phases  $\theta_{{\rm d}}(\vr_{\rm d})$ randomly chosen from $\{0,\pi\}$ or from $\{-\pi/2,\pi/2\}$, as indicated in the figure. The thin black lines denotes the ideal surface-state density of states without dislocation lines. The arrow denotes a Dirac delta function at zero energy. Data points denote an average over $5$ disorder
realizations and $80$ values of the crystal momentum $\kappa_x$.}
\end{figure}

We have also studied the cases in which the phases $\theta_{\rm d}(\vr_{\rm d})$ randomly fluctuate between the values $0$ and $\pi$, or between $\pi/2$ and $-\pi/2$. Results for transport properties and for the density of states are shown in Figs.\ \ref{fig:results_transport3} and \ref{fig:dos2}.

\end{document}